\documentclass[%
 reprint,
 amsmath,amssymb,
 aps,
 nofootinbib,
 onecolumn,
longbibliography,
preprintnumbers,
floatfix
]{revtex4-2}

\usepackage{graphicx}
\usepackage{dcolumn}
\usepackage{bm}
\usepackage{bbm}
\usepackage{xspace}
\usepackage{verbatim}
\usepackage{subcaption}
\DeclareCaptionJustification{justified}{\leftskip=0pt\rightskip=0pt\parfillskip=0ptplus1fil\relax}
\usepackage[normalem]{ulem}
\usepackage{xcolor}

\newcommand{\lea}{\Big\langle\!\!\Big\langle}
\newcommand{\rea}{\Big\rangle\!\!\Big\rangle}

\newcommand{\phistar}{\phi^{*}\,\!}
\newcommand{\Gphph}{ G_{\phi^{*} \phi} }
\newcommand{\Gss}{ G_{SS} }
\usepackage[
colorlinks=true,
linkcolor=blue,
citecolor=blue,
filecolor=blue,
urlcolor=blue,
]{hyperref}
\usepackage[capitalise]{cleveref}
\usepackage{pgfplots}
\newcommand{\SU}{\operatorname{SU}}

\renewcommand{\d}{\mathrm{d}}
\renewcommand{\i}{\mathrm{i}}

\usepackage{amsthm}
\theoremstyle{plain}

\newcommand{\<}{\langle}
\renewcommand{\>}{\rangle}

\newcommand{\su}{\mathfrak{su}}
\newcommand{\hl}[1]{{\color{green!50!black}{\bf hl}: #1}}

\newcommand{\LANL}{Theoretical Division, Los Alamos National Laboratory, Los Alamos, New Mexico, 87545}
\begin{document}

\preprint{LA-UR-25-20460}

\title{Universality of $SU(\infty)$ relaxation dynamics for $SU(n_f)$-symmetric spin-models.}

\author{Duff Neill}%
 \email{dneill@lanl.gov}
\author{Hanqing Liu}%
 \email{hanqing.liu@lanl.gov}
\affiliation{%
\LANL
}%
\date{\today}

\begin{abstract}
Spin-models, where the $N$ spins interact pairwise with a $SU(n_f)$ symmetry preserving hamiltonian, famously simplify in the large $n_f$, $N$ limits, as derived by Sachdev and Ye when exploring mean-field behavior of spin-glasses. We present numerical evidence that for a large class of models, the large $n_f$ limit is not necessary: the same dynamical equations can describe the relaxation processes at high temperatures for a set of classical models inspired from mean-field treatments of interacting dense neutrino gases, up to times set by the radius of convergence of the perturbation series for the correlation function. After a simple rescaling of time, the dynamics display a surprising universality, being identical for any value of $n_f$ as long as the rank of the coupling matrix is small. As a corollary of our results, we find that the direct interaction approximation originating from the study of stochastic flows in fluid turbulence should be thought of as only a short-time approximation for generic random coupling systems.
\end{abstract}
\maketitle


\section{Introduction}

A critical question in any physical system is how it relaxes back to equilibrium after a small perturbation, where such relaxation dynamics is typically captured by the correlation  function. One way physicists have made progress on this question is to consider all-to-all randomly coupled models of $N$-spins, where geometry does not matter, and the equations of motion often simplify in a dramatic fashion in the $1/N$ expansion. For a few examples both quantum and classical see Refs.~\cite{1961JMP.....2..124K,1994PhRvE..49.3990E,1993PhRvL..70.3339S,2016PhRvD..94j6002M}. Often to derive these simplified equations of motion, additional symmetry is assumed in the system allowing multiple large $N$ expansions, as in Ref.~\cite{1993PhRvL..70.3339S}. 

Neutrinos interacting in a dense supernova environment give an example of such an all-to-all coupled system \cite{Pastor:2001iu,Balantekin:2006tg}, however with dramatically different assumptions on the nature of the coupling matrix than typically assumed in the condensed matter literature. In particular, the coupling matrix for the neutrino system is rank deficient, with the rank of the matrix much less than the number of interacting spins. As shown in Ref.~\cite{Neill:2024klc} the resulting thermodynamics for such models do not display any spin-glass behavior. In this paper, we provide numerical evidence and heuristic arguments about the classical relaxation process for such $SU(n_f)$-symmetric spin systems: the short-time relaxation dynamics always behave as if $n_f=\infty$, so that the resulting dynamics is governed by the same set of equations made famous by the Sachdev-Ye-Kitaev (SYK) model for quantum chaos \cite{1993PhRvL..70.3339S,2016PhRvD..94j6002M}.\footnote{That such equations could result in classical processes was recently pointed out in Ref.~\cite{2023PhRvE.108e4132H}, but long appreciated in the fluid dynamics literature \cite{Kraichnan_1959,1961JMP.....2..124K,1981PhFl...24..615H,1994PhRvE..49.3990E}.} Moreover, we find the time-scale where we depart from the $SU(\infty)$ dynamics appears to be set by the radius of convergence of the Taylor series for the correlation function, a quantity that figures prominently in discussions of classical chaos, operator growth and complexity \cite{SIGETI1995136,Parker:2018yvk}, and recently a subject of an intense study in Ref.~\cite{Dodelson:2024atp} for specifically the SYK model.

In what follows, we will first introduce the spin-models of interest, and after a more precise statement of the conjecture, we will present our numerical evidence, and conclude.

\section{Dynamics of Spin-Models}
We consider a collection of classical spins interacting via the Heisenberg interaction, with each spin being a unit vector in 3 dimensions. We label the spins from $1$ to $N$, and we let $\Lambda=\{1,...,N\}$ be the set of all labels. The Hamiltonian of the basic spin model we wish to consider is:
\begin{align}
\label{eq:_dyn_cl_rotor_H}H_{\rm cl.}&=\sum_{\{i,j\}\subset \Lambda}  J_{ij}\vec{S}_i\cdot \vec{S}_j\,.
\end{align}
Our spins are unit vectors in $\mathbb{R}^{n_f^2-1}$, and thus have a natural mapping into the adjoint representation of $SU(n_f)$. We will assume that the couplings $J_{ij}$ will be randomly drawn from some distribution. The equations of motion that maintain the $SU(n_f)$ symmetry of the system can be derived from Eq.~\eqref{eq:_dyn_cl_rotor_H} as long as we modify our Poisson bracket to be:
\begin{align}\label{eq:pb}
    \{S_i^a,S_j^b\}_{\rm P.B.}&=\delta_{ij}\sum_{c=1}^{n_f^2-1}f^{abc}S_{i}^{c}\,,
\end{align}
with $f^{abc}$ being the structure constants of the $SU(n_f)$, and $\{\cdot,\cdot\}_{\rm P.B.}$ is the classical Poisson Bracket. This results in the classical equations of motion for the spin, the generalization of the Landau-Lifshitz equation for $SU(n_f)$ \cite{PhysRevB.104.104409}:
\begin{align}\label{eq:hamiltonian_eq_motion}
    \frac{d}{dt}S_i^a=\{S_i^a,H_{\rm cl.}\}=\sum_{j\in \Lambda}\sum_{c,b=1}^{n_f^2-1}J_{ij}f^{abc}S_i^{b} S^{c}_j\,.
\end{align}
Alternatively, similar to what is done when one wishes to formulate a coherent state path integral for spins, we can write down a Lagrangian formulation of the same dynamics, using the fact that the adjoint representation is the ``square'' of the fundamental representation:
\begin{align}\label{eq:spin-to-phi}
    S_i^{a}=\phistar_i\!\cdot\! T^{a}\!\cdot\!\phi_i\,.
\end{align}
$T^a$ are the generating matrices for the fundamental representation, and $\phi_i$ is a $n_f$-index complex spinor. Then the appropriate Lagrangian for the dual variables is, (see App. \ref{app:L}):\footnote{When we adopt such a lagrangian, the $\phi_i$ and $\phistar_i$ should be thought of as independent fields with independent equations of motion, whose specific norms are not conserved, but only the combination $\sum_{i}\phistar_i\cdot\phi_i$. This allows for an unconstrained equations of motion for our fields.}
\begin{align}\label{eq:L}
    L=\frac{1}{2i}\sum_{i\in\Lambda}\phistar_i\!\cdot\!\frac{d\phi_i}{dt}-\phi_i\!\cdot\! \frac{d\phistar_i}{dt}-\sum_{\{i,j\}\subset \Lambda}\sum_{a=1}^{n_f^2-1} J_{ij}(\phistar_i\!\cdot\! T^{a}\!\cdot\!\phi_i)(\phistar_j\!\cdot\! T^{a}\!\cdot\!\phi_j)\,.
\end{align}
It is straightforward to check that the resulting standard Lagrangian equations of motion are equivalent to Eq.~\eqref{eq:hamiltonian_eq_motion}. 

What we are concerned about is the behavior of the two-point auto-correlation function for the spin model at a specific inverse temperature $\beta$:
\begin{align}\label{eq:spin-spin-auto}
    \Gss(t)&=\frac{1}{N}\sum_{i\in\Lambda}\Theta(t)\lea \vec{S}_{i}(t)\cdot \vec{S}_i(0)\rea_{\beta}\,,\\
    \lim_{t\rightarrow 0^+}\Gss(t)&=1\,.
\end{align}
In short, we generate a spin configuration that is at inverse temperature $\beta$ as given by the Gibbs distribution, and evolve the spins forward using the classical equations of motion, calculating the overlap with the initial configuration. We write the double brackets to denote that we average over both the multiple configurations at finite temperature, and over the random realizations of the coupling matrix. We can define a similar correlation function for our dual variables:
\begin{align}
    \Gphph(t)&=\frac{1}{N}\sum_{i\in\Lambda}\Theta(t)\lea \phistar_{i}(t)\!\cdot\! \phi_i(0)\rea_{\beta}\,.
\end{align}

Using the $SU(n_f)$ relations:
\begin{align}        T^a_{ij}T^a_{k\ell}=\delta_{i\ell}\delta_{jk}-\frac{1}{n_f}\delta_{ij}\delta_{k\ell}\,,
\end{align}
we achieve:
\begin{align}
    \Gss(t)&=\frac{1}{N}\sum_{i\in\Lambda}\Theta(t)\lea (\phistar_{i}(t)\!\cdot\! \phi_i(0))(\phistar_{i}(0)\!\cdot\! \phi_i(t))\rea_{\beta}-\frac{1}{n_f}\frac{1}{N}\sum_{i\in\Lambda}\Theta(t)\lea (\phistar_{i}(t)\!\cdot\! \phi_i(t))(\phistar_{i}(0)\!\cdot\! \phi_i(0))\rea_{\beta}\,.
\end{align}
Note that $\phistar_{i}(t)\!\cdot\! \phi_i(t)$ is a conserved quantity, so that the last term is manifestly negative and a constant. In fact, given the normalization at time $t=0$ of $\Gss$, we find:
\begin{align}
\frac{n_f}{n_f-1}=\frac{1}{N}\sum_{i\in\Lambda}\lea(\phistar_{i}(0)\!\cdot\! \phi_i(0))^2\rea_{\beta}\,.
\end{align}

For finite $n_f$, we have simply recast the theory into new variables, which may or may not buy us anything except improved numerics \cite{PhysRevB.106.235154}. However, in Ref.~\cite{1993PhRvL..70.3339S}\footnote{Technically, they considered the corresponding quantum models to the above classical hamitlonian of Eq.~\eqref{eq:_dyn_cl_rotor_H}}, it was argued that if we take our spins $\vec{S}_i$ to appropriately transform in the $SU(n_f)$ symmetry group, and take $n_f\rightarrow\infty,N\rightarrow\infty$, we have the relations\footnote{Note that the way we take $n_f$ and $N$ to infinity will have profound effects on the phase diagram of the model, but at least at sufficiently high temperatures, we need not worry about this.}:
\begin{align}
\frac{1}{N}\sum_{i\in\Lambda}\lea \phistar_{i}(t)\!\cdot\! \phi_i(0)\rea_{\beta} &=\Gphph(t)+...\,,\\
    \Gss(t)&=|\Gphph(t)|^2\,.\label{eq:ss_to_phph}
\end{align}
 We have also assumed that we can replace the average square with the square of the average in the thermal and coupling ensembles in Eq.~\eqref{eq:ss_to_phph}, an assumption that should fail in a spin-glass phase (see Sec. \ref{sec:spin_glass}). Lastly the $\phi-\phi$ correlation  function will satisfy:
\begin{align}\label{eq:syk_two_point}
\frac{d}{dt}\Gphph(t) &= \delta(t)+ \int\displaylimits_{0}^{t} ds \Sigma(s) \Gphph(t-s)\,,\\
\Sigma(t) &= -J^2\Big( \Gphph(t) \Big)^3\,.
\end{align}
The last line only holds strictly at infinite temperature, and $J$ is related to the variance of the random couplings $J_{ij}$, with $\Sigma(t)$ the self-energy calculated from resumming melonic diagrams. At finite temperatures, the self-energy takes a slightly more involved form, as easily seen by investigating the so-called Sachdev-Ye-Kitaev model that generalizes these results to interacting Majorana fermions, where the self-energy must account for the fermion statistics~\cite{2016PhRvD..94j6002M}. We note that these results were derived with the specific assumption that the coupling between any pair of spins is gaussian distributed, an assumption which we will relax.

Such closure equations for the two-point function have a longer history, dating back to early studies of fluid turbulence \cite{Kraichnan_1959}, where it was called the direct-interaction approximation (DIA). The closures can be formulated at the level of the equations of motion themselves, so that the set of equations need not a lagrangian or hamiltonian description, see Refs.~\cite{1981PhFl...24..615H,2023PhRvE.108e4132H}\footnote{See Ref.~\cite{2023PhRvE.108e4132H} for a clear discussion for how the DIA connects with the path-integral formulation typical of the spin-glass literature, as well as a review for how the older turbulence literature derived the closures without such a formulation.}. In general, if we have a non-linearity in the relevant variables of polynomial order $q$ in the action, or $q-1$ in the equations of motion, the closure reads:
\begin{align}\label{eq:DIA_various_q}
    \frac{d}{dt}G(t)=\delta(t)-\alpha^2\int\displaylimits_{0}^{t}ds(G(s))^{q-1}G(t-s)\,.
\end{align}
The DIA closure has been derived under much milder assumptions about the underlying nature of the random coupling, see Ref.~\cite{1961JMP.....2..124K}, assuming only a mean and a variance for the coupling distribution, not the full probability distribution. Since our coupling distribution will not in general conform to the random Gaussian ensemble, we will henceforth call such closures the DIA.

Our specific conjecture is that the class of spin-models considered in Ref.~\cite{Neill:2024klc}, where the rank of the random coupling matrix is much less than the number of spins, satisfies (at high temperatures) the same set of DIA closure equations as the original Sachdev-Ye models when $N\rightarrow\infty$. This is \emph{regardless} of the value of $n_f$, as long as we appropriately rescale the effective coupling in the DIA equations, according to a value that can be calculated from the initial thermodynamic state. That is, at any value  of $n_f$, down to temperatures determined by the largest eigenvalue of the coupling matrix (see \cite{Neill:2024klc}), and up to a time $t_{th}$, the system behaves as if it is in the $n_{f}=\infty$ symmetry group:
\begin{align}\label{eq:pre_DIA} 
    \Gss(t)&\approx(\Gphph(t))^2, \text{ for }t\leq t_{th}\,,\\
\label{eq:DIA}    \frac{d}{dt}\Gphph(t) &= \delta(t)- \Big(\alpha(\beta)\Big)^2\int\displaylimits_{0}^{t} ds \Big( \Gphph(s) \Big)^3 \Gphph(t-s)\,,\\
\label{eq:eff_coupl}   \Big(\alpha(\beta)\Big)^2 &= \frac{2}{3N}\lea \sum_{\{i,j\}\subset\Lambda} \big(J_{ij}\big)^2\vec{S}_i\cdot\vec{S}_j \rea_{\beta}\sim O\Big( \frac{1}{N}\text{rank} J\Big)\,, 
\\
   \alpha(\beta)t_{th}&\approx 1.78\,.
\end{align}
The effective coupling $\alpha(\beta)$ is determined from expanding out the equations of motion for $\vec{S}_i$ (Eq.~\eqref{eq:hamiltonian_eq_motion}) to quadratic order in the time-evolution for the spin-spin autocorrelation function of Eq.~\eqref{eq:spin-spin-auto}, and evaluating at $t=0$. We then demand the curvatures for the $\phi$-auto-correlation function to match the spin's auto-correlation function. The time-scale of $\sim1.78$ where we expect the results to diverge is set by the radius of convergence of the $\Gphph$ implied by the DIA Eq.~\eqref{eq:DIA}, which we determine numerically by calculating about 30 of the terms of the taylor series, and then estimating based on the ratio test. 

Our conjecture takes inspiration from the observation made in Refs.~\cite{1998JPhA...31.9871G,PhysRevB.104.104409}, that the exact symmetry group we use to take a classical limit of quantum spins is often ambiguous. So then if it is ambiguous, it may as well be $n_f=\infty$, and the ambiguity is only resolved by the dynamics at sufficiently late times when the system relaxes back to thermal equilibrium. Another way we can heuristically derive these results is to consider that any $SU(n_f)$ group can be embedded as a subgroup of $SU(\infty)$, and so when we write the relation between the spinors and adjoint representation in Eq.~\eqref{eq:spin-to-phi}, we can take the $\phi$ to live in $SU(\infty)$, and relate them to our spins $\vec{S}_i$ through a set of generators $T^a$ that constitute some $SU(n_f)$ subgroup. Again, at sufficiently late-times in the relaxation process, the true dynamics is constrained to have only symmetry with respect to $SU(n_f)$, and we no longer expect our DIA equation to hold. 

Finally, we note that we could have attempted to formulate the DIA directly on the equations of motion for the spin variables $\vec{S}_{i}$ in Eq.~\eqref{eq:hamiltonian_eq_motion}, following the same set of steps as Ref.~\cite{1981PhFl...24..615H}. This would result in:
\begin{align}\label{eq:q2_DIA}
    \frac{d}{dt}\Gss=_{?}\delta(t)-\alpha^2\int\displaylimits_{0}^{t}ds \Gss(s)\Gss(t-s)\,.
\end{align}
That is the DIA with $q=2$. As we will see below, the integration of the hamiltonian equations of motion manifestly favors Eqs.~\eqref{eq:pre_DIA} to \eqref{eq:eff_coupl}, and not Eq.~\eqref{eq:q2_DIA}.

\section{Numerical Results}
In what follows, we will consider two distinct models, both discussed in Ref.~\cite{Neill:2024klc}:
\begin{align}\label{eq:rom_def}
&\text{\bf{Random Orthogonal Model:} }J_{ij}=\sum_{k=1}^{M}O_{ik}O_{jk}e_k, e_{k}=-2+4\frac{k}{M}\,,\\ 
&\qquad\qquad\qquad\nonumber O_{ij}\text{ random orthogonal }  N\times N\text{ matrix, selected with Haar measure. }\\
&\text{\bf{Neutrino Model:} }J_{ij}=\frac{1}{N}(1-\vec{v}_i\cdot\vec{v}_j)\,, \text{ with } \vec{v}_i \text{ uniformily distributed on the unit sphere. }
\end{align}
Our neutrino model is naturally rank deficient, having 4 non-trivial eigenvalues in the large $N$ limit. The random orthogonal model allows us to easily control the rank of the coupling matrix. As shown in Ref.~\cite{Neill:2024klc}, these models can be solved in the high temperature limit, with a phase transition at a temperature related to the largest eigenvalue of the coupling matrix. We will scan over these high temperatures, and compare the short-time decay of the two-point function to the DIA results. 

Our numerics for the spin-models are calculated as follows: we generate an instance of the coupling matrix according to the desired distribution, then we use a simulated annealing heat bath algorithm to generate an initial configuration at a desired temperature scale. We then evolve the initial conditions via the hamiltonian classical equations of motion given in Eq.~\eqref{eq:hamiltonian_eq_motion} to a time scale on the order of $10\times \sqrt{N/(\text{rank} J)}$, an estimate based on Eq.~\eqref{eq:eff_coupl}. At regular intervals, we record the average auto-correlation of the spins with themselves, giving us the function $\Gss$. 

We then rescale the time on all our spin auto-correlation functions to agree with the determination of $(\Gphph)^2$ at its half-max from the DIA in Eq.~\eqref{eq:DIA} setting $\alpha(\beta)=1$. We could directly use the relation in Eq.~\eqref{eq:eff_coupl} to do this, but instead we determine the rescaling factor directly from the time-evolved results. This allows us to compare to the effective coupling from the time evolved results to the thermodynamic calculation in Eq.~\eqref{eq:eff_coupl}. \hl{See \cref{fig:rom_eff_coupl}} Concretely, to determine the effective coupling in the DIA equation from the time-evolved $\Gss$, we define the following time scales as:
\begin{align}
\frac{1}{2}&=\Gss(t^{SS}_{\frac{1}{2}})\,,\\
\frac{1}{2}&=
\Big(\Gphph(t^{\rm DIA}_{\frac{1}{2}})\Big)^2\Big|_{\alpha=1}\,.
\end{align}
We note that at this value, the Taylor series at quadratic order for $\Gphph$ and $\Gss$ do not provide a good approximation of these functions. We then can determine the effective coupling in the DIA from the time-evolution as: 
\begin{align}\label{eq:half_max_time}
\alpha = \frac{t^{\rm DIA}_{\frac{1}{2}}}{t^{SS}_{\frac{1}{2}}}\,.
\end{align}
Rather than solve the DIA with each new coupling thus found, we rescale the time argument of $\Gss$ to compare to the DIA equation with $\alpha=1$ in all figures.

\subsection{Rank Deficient Random Orthogonal Models}
\begin{figure}
    \centering
    \includegraphics[scale=0.56]{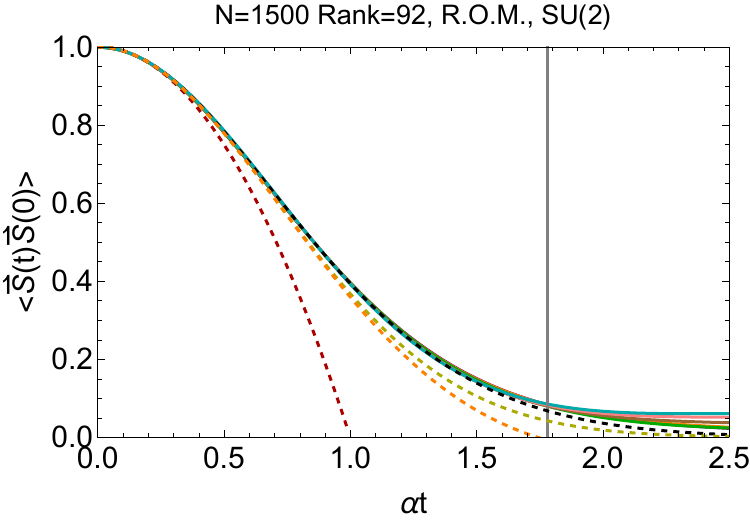}
    \includegraphics[scale=0.56]{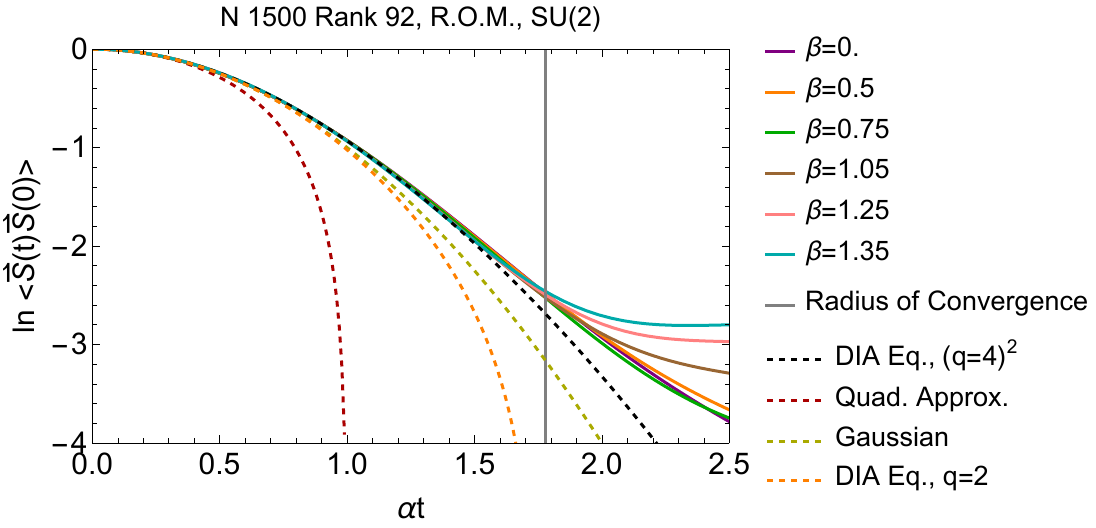}
        \caption{We plot the spin-spin auto-correlation function (Eq.~\eqref{eq:spin-spin-auto}) for the random orthogonal model with $SU(2)$ symmetry with $N=1500$ and rank $J$=92, on both a linear scale (left) and a log scale (right). We compare against the DIA for $q=4$ \eqref{eq:pre_DIA}\&\eqref{eq:DIA} and $q=2$ \eqref{eq:q2_DIA}, the quadratic approximation to the taylor series, as well as a gaussian with the same curvature. Up to times set by the radius of convergence of the DIA taylor series, we see excellent agreement with the DIA results with $q=4$ (\eqref{eq:pre_DIA}\&\eqref{eq:DIA}) for all temperatures. We note the statistical fluctuations on these curves are too small to represent.}
    \label{fig:1500_92_SU2}
\end{figure}
\begin{figure}
    \centering
    \includegraphics[scale=0.56]{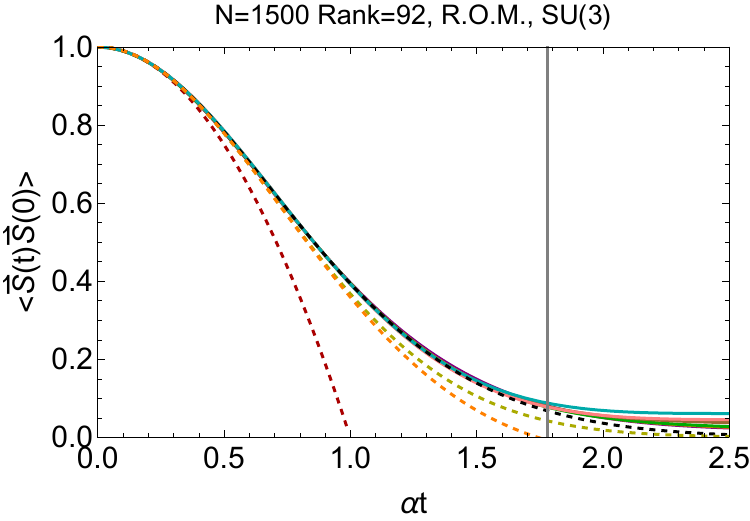}
    \includegraphics[scale=0.56]{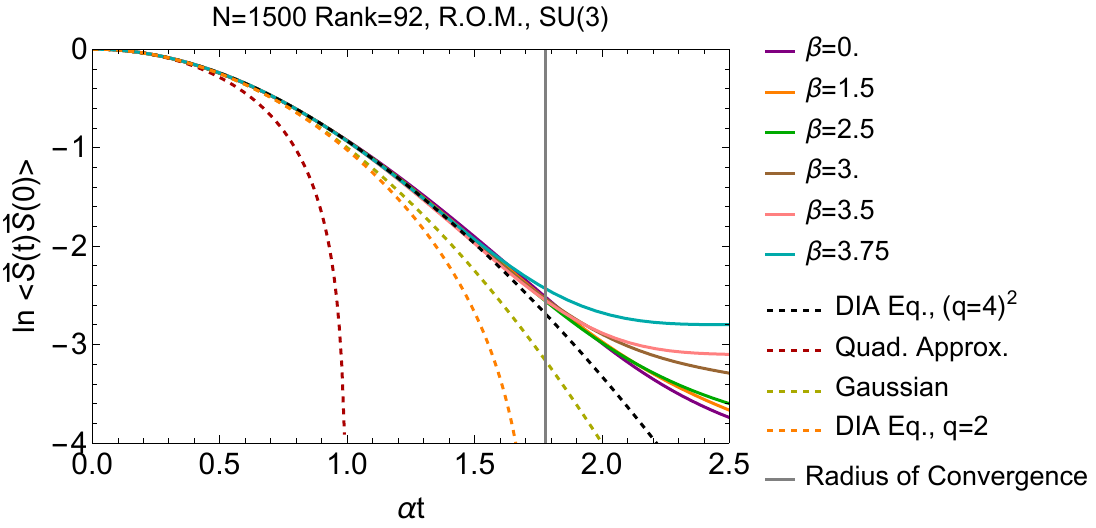}
    \caption{We plot the spin-spin auto-correlation function (Eq.~\eqref{eq:spin-spin-auto}) for the random orthogonal model with $SU(3)$ symmetry with $N=1500$ and rank $J$=92, on both a linear scale (left) and a log scale (right). We compare against the DIA for $q=4$ \eqref{eq:pre_DIA}\&\eqref{eq:DIA} and $q=2$ \eqref{eq:q2_DIA}, the quadratic approximation to the taylor series, as well as a gaussian with the same curvature. Up to times set by the radius of convergence of the DIA taylor series, we see excellent agreement with the DIA results with $q=4$ (\eqref{eq:pre_DIA}\&\eqref{eq:DIA}) for all temperatures. We note the statistical fluctuations on these curves are too small to represent.}
    \label{fig:1500_92_SU3}
\end{figure}
\begin{figure}
    \centering
    \includegraphics[scale=0.58]{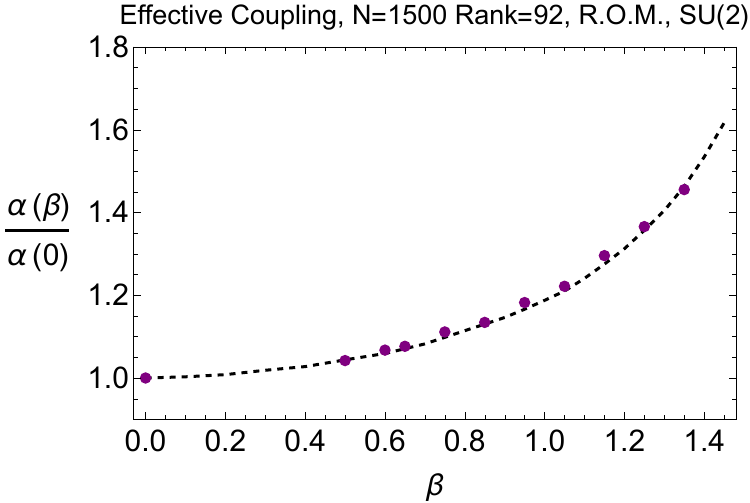}
    \includegraphics[scale=0.58]{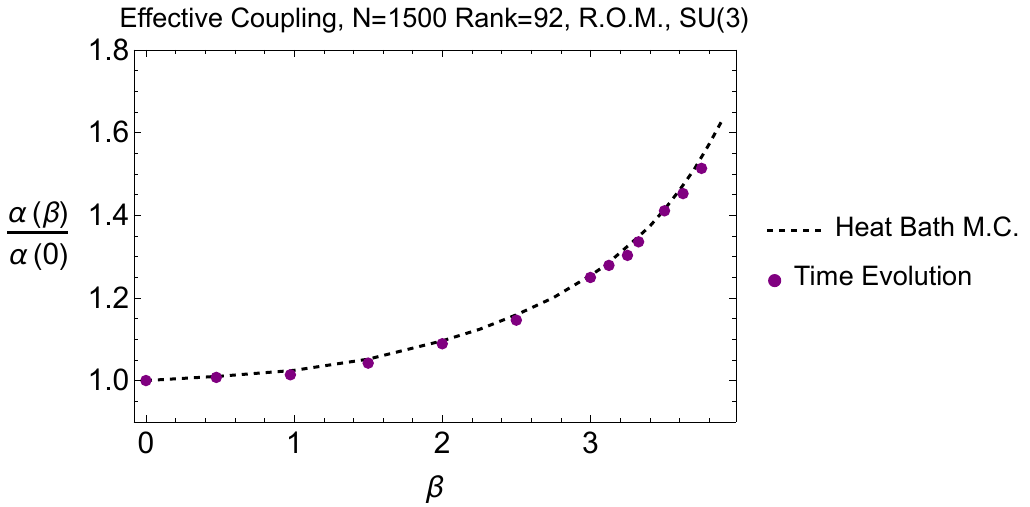}
    \caption{We plot the effective coupling in the DIA equations for the both the SU(2) and SU(3) random orthogonal models. The dots correspond to the determination from rescaling the time-evolved solution to agree with the DIA \eqref{eq:pre_DIA}\&\eqref{eq:DIA} at half-max, Eq.~\eqref{eq:half_max_time}, while the dashed lines are the calculation of Eq.~\eqref{eq:eff_coupl} from a thermal calculation using a heat-bath monte carlo sampling. We normalize the effective coupling by the result at infinite temperature.}
    \label{fig:rom_eff_coupl}
\end{figure}

We give example the two-point function for $N=1500$, for a coupling matrix defined as in Eq.~\eqref{eq:rom_def} with rank$(J_{ij})=92$ for spins in both $SU(2)$ and $SU(3)$. For spins taking values as unit vectors in $d_{r}$ dimensions, the results of Ref.~\cite{Neill:2024klc} predicts we have a phase-transition at $\beta=\frac{d_{r}}{2}$. For the $SU(2)$ model this is $\beta=3/2$, while for $SU(3)$, this is $\beta=4$. As we lower temperature, the effective coupling $\alpha(\beta)$ generically increases, eventually diverging near the phase-transition, as can be seen in Fig.~\ref{fig:rom_eff_coupl}. 

\begin{figure}
    \centering
    \includegraphics[scale=0.56]{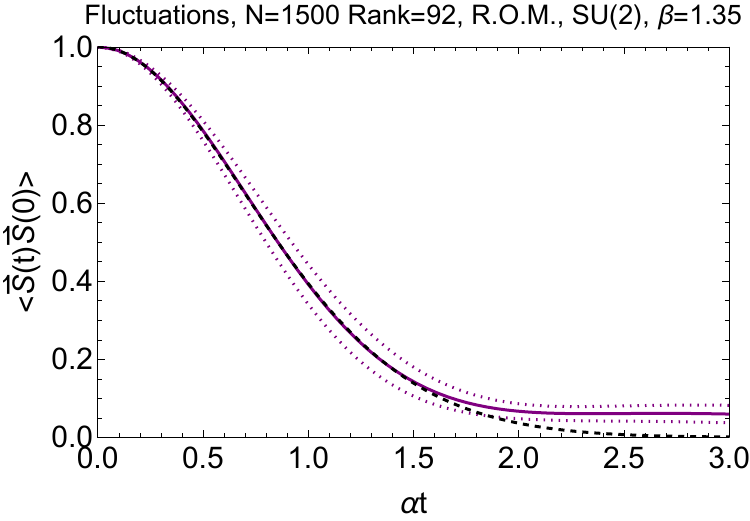}
    \includegraphics[scale=0.56]{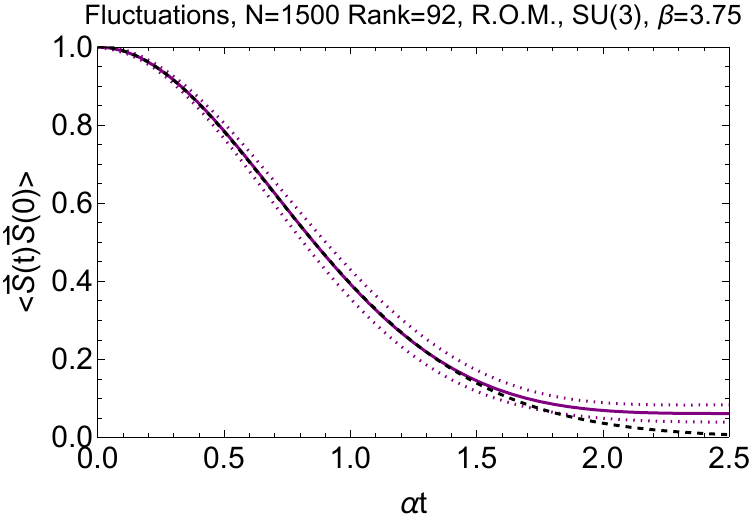}
    \caption{We plot the average spin-spin auto-correlation (Eq.~\eqref{eq:spin-spin-auto}, solid purple) function  and its root-mean square fluctuations (dotted purple) for the random orthogonal model with either $SU(2)$ (left) or $SU(3)$ symmetry with $N=1500$ and rank $J$=92. We take curves at temperatures close to the predicted phase transition in these models. We compare against the DIA \eqref{eq:pre_DIA}\&\eqref{eq:DIA}, the black dashed curve. We note that the root-mean-squared shown here should not be interpreted as the statistical error on the average auto-correlation function, but its genuine fluctuations under the thermal and coupling ensemble averages.
    }
    \label{fig:ROM_fluct}
\end{figure}

In Fig.~\ref{fig:1500_92_SU2}, we show the results for $SU(2)$ dynamics, while in Fig.~\ref{fig:1500_92_SU3} we show the results for $SU(3)$, both in linear and log scales. We see excellent agreement with the DIA Eq.~\eqref{eq:DIA} result up until the radius of convergence, where the two-point function at various temperatures does what it will. Further, in Fig.~\ref{fig:ROM_fluct}, we compare near the phase transition for both models the two-point functions and their typical root-mean-square fluctuations, again seeing excellent agreement with the DIA result. We note that the root-mean-squared shown there should not be interpreted as the statistical error on the average auto-correlation function, but its genuine fluctuations under the thermal and coupling ensemble averages.

\subsection{Neutrino Models}
\begin{figure}
    \centering
    \includegraphics[scale=0.56]{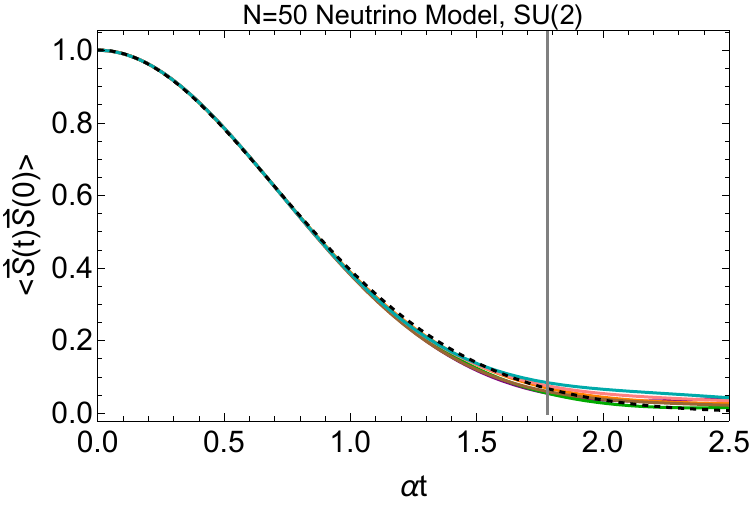}
    \includegraphics[scale=0.56]{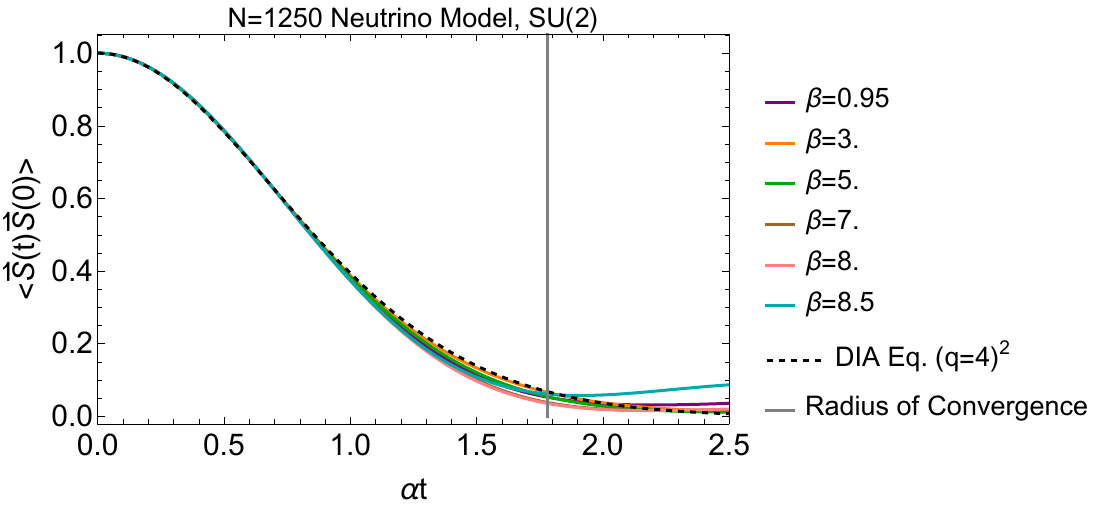}
    \caption{We plot the spin-spin auto-correlation function (Eq.~\eqref{eq:spin-spin-auto}) for the neutrino coupling model with $SU(2)$ symmetry with $N=50$ and $N=1250$. We compare against the DIA \eqref{eq:pre_DIA}\&\eqref{eq:DIA}. We see larger departures from the DIA, but generally up to times set by the radius of convergence of the DIA taylor series, the DIA results describe the average auto-correlation function for all temperatures up to the phase-transition. We note the statistical fluctuations on these curves are too small to represent.}
    \label{fig:neutrino}
\end{figure}
\begin{figure}
    \centering
    \includegraphics[scale=0.56]{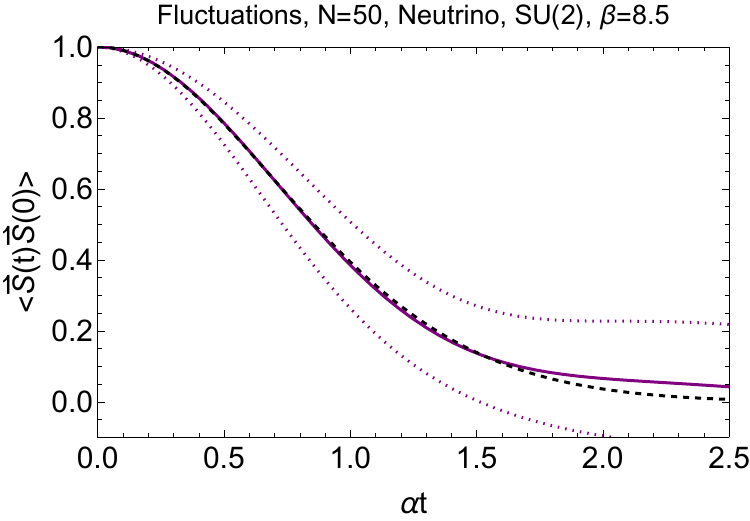}
    \includegraphics[scale=0.56]{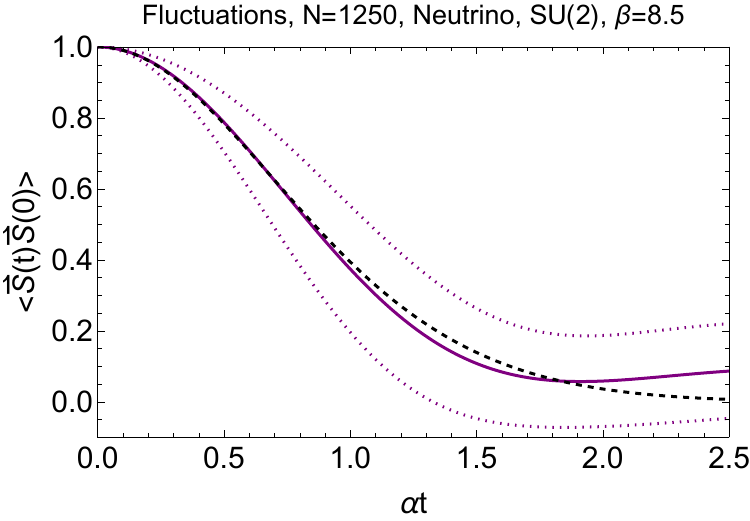}
    \caption{We plot the average spin-spin auto-correlation (Eq.~\eqref{eq:spin-spin-auto}, solid purple) function and its root-mean square fluctuations (dotted purple) for the neutrino coupling model with SU(2) symmetry, for $N=50$ and $N=1250$. The fluctuations are much large compared to the random orthogonal model case. We take curves at temperatures close to the predicted phase transition in these models. We compare against the DIA \eqref{eq:pre_DIA}\&\eqref{eq:DIA}, the black dashed curve. We note that the root-mean-squared shown here should not be interpreted as the statistical error on the average auto-correlation function, but its genuine fluctuations under the thermal and coupling ensemble averages.}
    \label{fig:neutrino_fluct}
\end{figure}
In Fig.~\ref{fig:neutrino}, we show the results for $SU(2)$ dynamics in the neutrino coupling model. We see good agreement with the DIA Eq.~\eqref{eq:DIA} result up until the radius of convergence, where the two-point function at various temperatures does what it will, but slightly earlier than the random orthogonal models considered before. The phase-transition for these models is at a temperature of $\beta = 9$. Further, in Fig.~\ref{fig:neutrino_fluct}, we compare near the phase transition for the model the two-point functions and their typical root-mean-square fluctuations, again seeing excellent agreement with the DIA result. We note that the root-mean-squared shown there should not be interpreted as the statistical error on the average auto-correlation function, but its genuine fluctuations under the thermal and coupling ensemble averages.

\subsection{Full Rank Random Orthogonal Models}\label{sec:spin_glass}
 Qualitatively different behavior emerges in the random orthogonal model with a full rank coupling matrix. In Ref.~\cite{Neill:2024klc}, it was noted at sufficiently low-temperatures, a heat-bath algorithm failed to decorrelate successive samples, indicating a spin-glass phase. While we made no attempt to find the spin-glass phase-transition temperature, the lack of decorrelation only became readily apparent when $\beta>3$. Similar spin-glasses to the one we encounter here have been studied in the literature, the so-called p-spin glass models, see Refs.~\cite{PhysRevLett.71.173,PhysRevB.59.915}, albeit with a different coupling structure, though classically a similar set of mean-field equations were obtained. 
 
\begin{figure}
    \centering
    \includegraphics[scale=0.56]{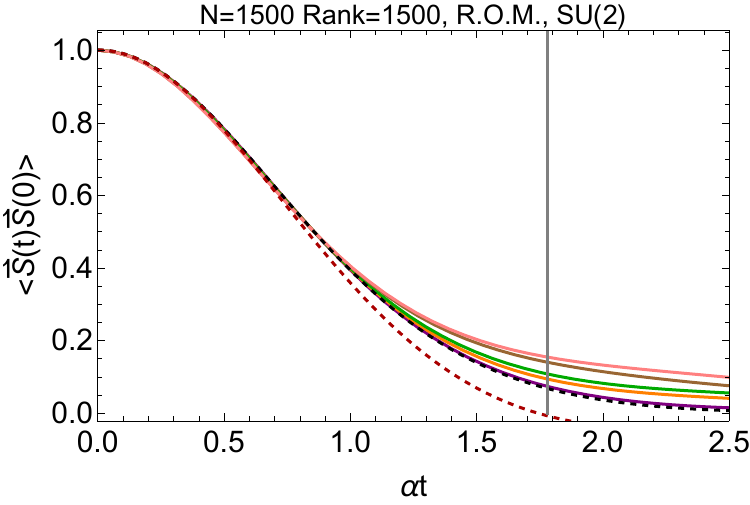}
    \includegraphics[scale=0.56]{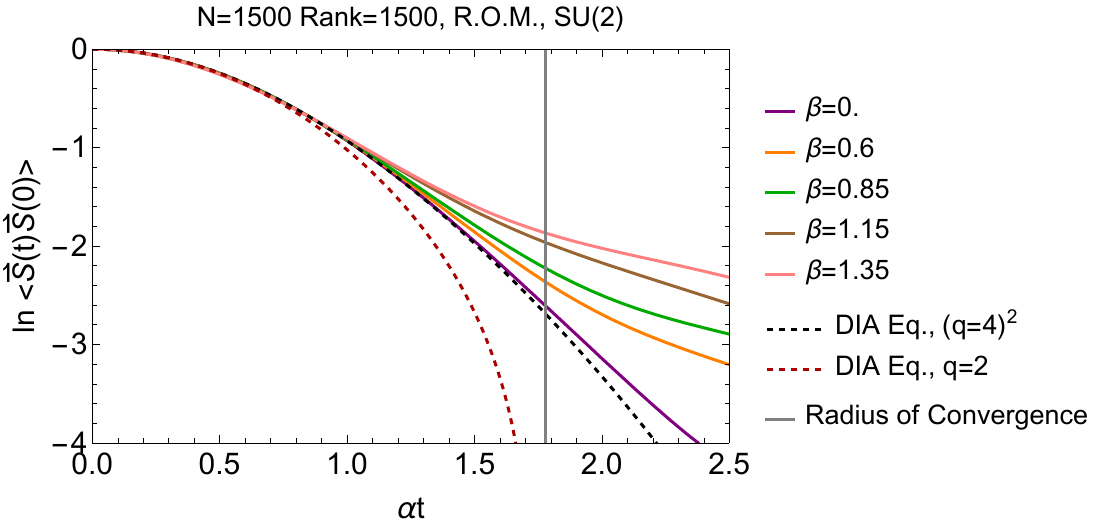}
    \caption{We plot the spin-spin auto-correlation function for the random orthogonal model with $SU(2)$ symmetry with $N=1500$ and rank $J$=1500, on both a linear scale (left) and a log scale (right). We compare against the DIA for $q=4$ (\eqref{eq:pre_DIA}\&\eqref{eq:DIA}) and $q=2$ (\eqref{eq:q2_DIA}).  We note the statistical fluctuations on these curves are too small to represent.}
    \label{fig:1500_1500_SU2}
\end{figure}

 As we lower the temperature and examine the two-point spin auto-correlation function, we can see in Fig.~\ref{fig:1500_1500_SU2} even without being close to the region where spin-glass behavior could be observed in the monte-carlo heat-bath, the models begin to depart steadily from the DIA \eqref{eq:pre_DIA}\&\eqref{eq:DIA}, though this model most closely follows it at infinite temperatures.

\begin{figure}
    \centering
    \includegraphics[scale=0.56]{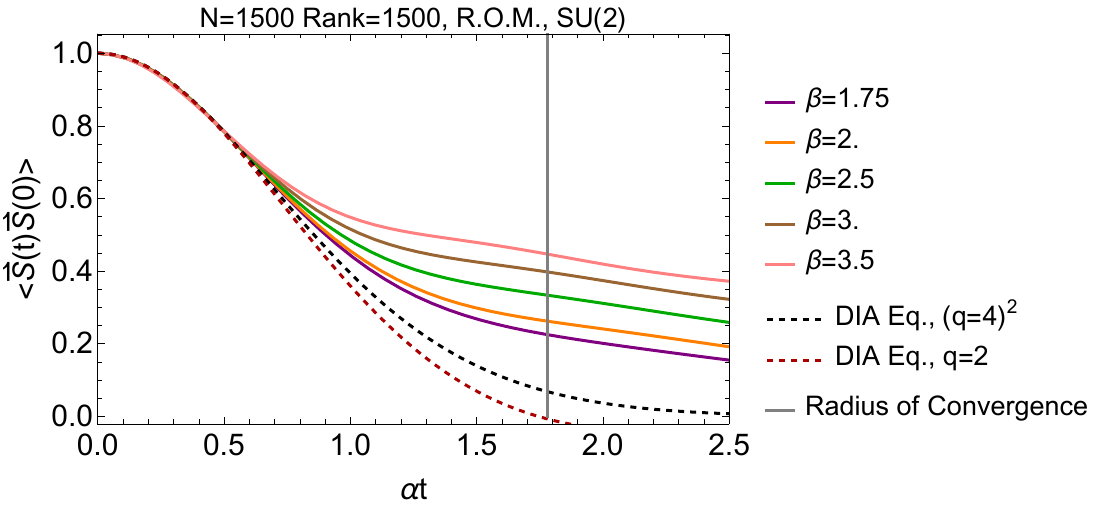}
    \caption{We plot the spin-spin auto-correlation function for the random orthogonal model with $SU(2)$ symmetry with $N=1500$ and rank $J$=1500, on both a linear scale for temperatures where spin-glass behavior can be observed. We compare against the DIA for $q=4$ (\eqref{eq:pre_DIA}\&\eqref{eq:DIA}) and $q=2$ (\eqref{eq:q2_DIA}). We note the statistical fluctuations on these curves are too small to represent.}
    \label{fig:1500_1500_SU2_lowtemp}
\end{figure}

 In Fig.~\ref{fig:1500_1500_SU2_lowtemp} we continue to see the same behavior as we lower the temperature further still, with ever more dramatic departures from the DIA results. Indeed, the rescaling of the time could no longer be determined from the half-max of the correlation functions, and we used a value of $0.8$. 

That we should see such departures from the DIA results is perhaps not surprising, as when we compare against them we are switching the average of the square of the $\phi-\phi$ correlation with the square of the average.
In spin glasses phases, this is an invalid procedure, at the heart of the difference between an annealled calculation versus a quenched. Indeed, in Ref.~\cite{PhysRevB.59.915} this is manifested by a non-trivial coupling between replicas in the mean-field equations. What is surprising is that even at temperatures where we do not yet expect spin glass behavior based on numerical evidence from the heat-bath M.C. decorrelates, we see departures from the DIA essentially at all temperatures below infinite, with a noticeable tail in the auto-correlation function.


\section{Radius Of Convergence}
\begin{figure}
    \centering
    \includegraphics[scale=0.6]{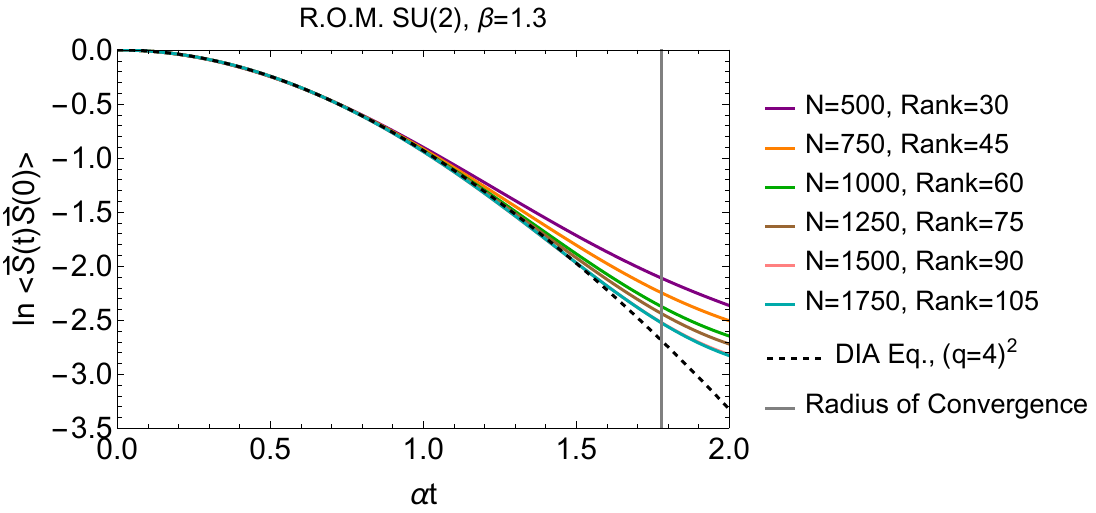}
    \caption{We plot the average spin-spin auto-correlation function for the random orthogonal model for various $N$, with the rank of the coupling matrix $J_{ij}$ linearly related to $N$, at a fixed temperature $\beta=1.3$. By eye, it is difficult to distinguish the $N=1500$ and $N=1750$ curves.}\label{fig:var_N}
\end{figure}

\begin{figure}
    \centering
    \includegraphics[scale=0.6]{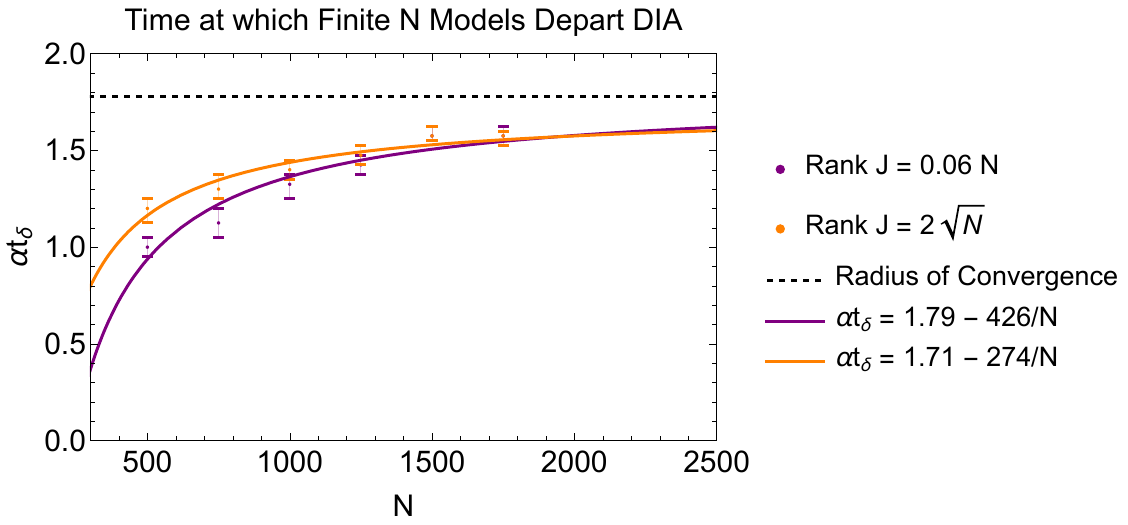}
    \caption{We plot the determination of $\alpha(\beta)t_{\delta}$ with $\delta = 0.03$, for the case where the rank of the coupling matrix $J_{ij}$ is linearly related to $N$ (purple), and varies as a square root of $N$ (orange), at a fixed temperature $\beta=1.3$.  The uncertainty reflects the statistical uncertainty of the $\Gss$ auto-correlation function. Different choices of $\delta$ result in different asymptotic values of $\alpha t_{\delta}$.}\label{fig:alpha_t_delta}
\end{figure}

Finally, we numerically explore whether or not the models depart from the DIA at $\alpha(\beta) t_{th}\approx 1.78$, the radius of convergence of the perturbation series for the DIA. Given a specific $N$, we define the time $t_{\delta}$ when we observe a departure from the DIA at a given magnitude $\delta$:
\begin{align}
    \delta = \frac{|(\Gphph( t_{\delta}))^2-\Gss( t_{\delta})|}{\Gss(t_{\delta})},
\end{align}
where we are careful to tune the effective coupling in the DIA Eq. for $\Gphph$ to match $\Gss$. We expect:
\begin{align}
    \lim_{\delta\rightarrow 0}\lim_{N\rightarrow \infty}\alpha(\beta) t_{\delta}=\alpha(\beta) t_{th}\,.
\end{align}
To explore this question, we generated for the $SU(2)$ random orthogonal model two families of curves where we take the large $N$ limit, with differing choices of how we change the rank of the coupling matrix, while keeping the temperature fixed. That is, we take the rank of the coupling matrix $J_{ij}$ to be either rank$J = 0.06N$ or rank$J \sim 2\sqrt{N}$, and compute the autocorrelation function at $N=500,750,1000,1250,1500,$ and $1750$. Fig.\ref{fig:var_N} plots the autocorrelation function when taking the rank of the coupling matrix to be a linear function of $N$, and Fig.\ref{fig:alpha_t_delta} gives the results for $\alpha t_{\delta}$ as a function of $N$, with a very naive extrapolation to infinity, where we fix $\delta=0.03$, and find the corresponding $\alpha t_{\delta}$. Given our numerical and statistical precision, we can estimate that:
\begin{align}
    \lim_{\delta\rightarrow 0}\lim_{N\rightarrow \infty}\alpha(\beta) t_{\delta}=1.8\pm 0.1\,.\\
    \lim_{\delta\rightarrow 0}\lim_{N\rightarrow \infty}\alpha(\beta) t_{\delta}=1.7\pm 0.1\,.
\end{align}
This is broadly consistent with $\alpha(\beta) t_{th}\approx 1.78$, and the uncertainity is estimated both by taking the envelope from both the statistical uncertainities in the auto-correlation functions, as well as varying the value of $\delta$ in the interval $[0.03,0.1]$. We cannot take $\delta$ to be much smaller than $0.03$, as we then begin to be limited by the numerical and statistical uncertainities in the determination of the auto-correlation function.

\section{Conclusions}

We have given numerical evidence and a heuristic argument for how SU($n_f$) spin systems exhibit remarkable universality in their relaxation dynamics. The identical DIA or mean-field equations one would derive for SU($\infty$) case apply to the classical dynamics of all-to-all spin systems for any $n_f$, up to times set by the radius of convergence for the taylor series of the DIA, as long as the coupling matrix is sufficiently simple (i.e., with a rank much less than the number of spins). One important lesson for the DIA, is that it must be understood as essentially a short-time approximation.

In the euclidean path integral formulation of these spin systems, one naturally uses the lagrangian \eqref{eq:L} for the action, since the standard lagrangian equations of motion are only consistent with the hamiltonian equations when one uses a modified poisson bracket.
From the derivations of the closure for the two-point function found in the fluid turbulence literature, one does not need a path integral, and only performs statistical averages for the coupling matrix on the equations of motion themselves. Naively, this leads to an ambiguity, whether one can postulate a DIA closure for the hamiltonian equations, or is it more correct to use the lagrangian equations for the dual variables? The choice leads to distinctly different short-time behavior, with the closure resulting from the lagrangian equations (\eqref{eq:pre_DIA}\&\eqref{eq:DIA}) clearly being favored by the direct numerical integration of the hamiltonian equation \eqref{eq:hamiltonian_eq_motion}. The distinguishing feature between the two formulations, as reviewed in the App.~\ref{app:H}, is that the $\phi$ variables can be taken to have canonical poisson brackets, that is, they are linearly related to standard flat space coordinates and momenta, but the spin variables $S$ must have a poisson bracket dictated by the Lie algrebra.

It would be interesting to explore in more depth the DIA as a means to close the neutrino quantum evolution in dense environments. We have only studied here the simplest reduction of these models to a SU($n_f$)-symmetric all-to-all interacting spin-model. It has long been debated what are the time-scales involved in these neutrino systems, Refs.~\cite{Friedland:2003eh,Bell:2003mg,Friedland:2006ke,Fiorillo:2024pns,Bhaskar:2024myw}. Often, the models considered in these cases have integrable dynamics, but for us, all of our models are chaotic and non-integrable. For the auto-correlation function for SU$(n_f)$-symmetric interactions, we find two time-scales: the dynamical time scale set by $\alpha(\beta)$ in Eq.~\eqref{eq:eff_coupl}, and the time for the onset of non-universal behavior, set by the radius of convergence for the perturbation series arising in Eq.~\eqref{eq:DIA}. True supernova neutrino interactions will have the SU$(n_f)$ symmetry broken by matter backgrounds and mass-induced vacuum oscillations, and further one must consider non-homogeneous distributions of neutrinos, modifying the time derivative to have an advective spatial component. The persistent questions that has arisen in the study of this problem is the boundary between when one can use the quantum kinetic equations (Refs.~\cite{Sigl:1993ctk,Vlasenko:2013fja}), their suitable classical counterparts, or the full quantum dynamics~\cite{Cervia:2019res,Martin:2021bri,Martin:2023gbo,Cirigliano:2024pnm}, and whether $n_f=2$ is sufficient for determining lessons on these boundaries~\cite{Turro:2024shh,Siwach:2024jet}. For our simplified models, we would again want to know whether the quantum dynamics follow the same closure equations for the two-point function.

Finally, we comment on how our conjecture can break down. First possibility is that the agreement observed between the numerical integration of the equations of motion and the DIA up until the radius of convergence of the Taylor expansion of the DIA is ephemeral: it could be that as $N\rightarrow\infty$, we get agreement at all times for any temperature above the phase-transition discussed in Ref.~\cite{Neill:2024klc}. Odd though this possibility may be, it would not be a terrible way to fail. Alternatively, the region where the DIA predicts the correlation function could be unrelated to the radius of convergence, it simply is another number. Another possibility is that as $N$ increases, we simply have a change in the behavior of the two-point function, and our conjecture is wrong. This seems the more unlikely possibility. 

The fact that the radius of convergence appears to play a key role in departures from the $SU(\infty)$ behavior ultimately must be related to ergodicity of the evolution. The chaotic time dynamics will  eventually explore the whole manifold available to it, and the difference between $SU(\infty)$ and $SU(n_f)$ will be manifested. At short-times, the evolution does not know what manifold it lives on, and so assumes a universal character. Presence of a spin-glass phase would break the ergodicity in a precise sense: it takes dramatically less time for the system to explore the available phase-space, so that departures from universal behavior come sooner and sooner as the temperature is lowered.

\section{Acknowledgements}
We'd like to thank Scott Lawrence, Cristian Batista, Joshua Martin, Alessandro Roggero, and Kipton Barros. Finally, we would like to thank M. Dodelson for answering a question concerning~\cite{Dodelson:2024atp}. This work was supported by the Quantum Science Center (QSC), a National Quantum Information Science Research Center of the U.S. Department of Energy (DOE) and by the U.S. Department of Energy, Office of Science, Office of Nuclear Physics (NP) contract DE-AC52-06NA25396.  

\bibliography{bib}

\appendix

\section{Hamiltonian formalism of classical spin dynamics}\label{app:H}
In this appendix, we give a self-contained review of the Hamiltonian dynamics of a classical $\SU(n_f)$ spin system. A natural way to obtain the classical limit of the quantum spin system in the fundamental representation\footnote{While the local Hilbert space in general can be any representation of $\SU(n_f)$ and the treatment is similar, here for simplicity, we only focus on the case considered in the main text, namely fundamental representation of $\SU(n_f)$.} of $\SU(n_f)$ is to restrict the full Hilbert space to product states of single-spin coherent states. Each single spin lives on the complex projective space $\mathbb{CP}^{n_f - 1}$. For a system of $N$ spins, the classical phase space is therefore
\begin{align}
  \underbrace{\mathbb{CP}^{n_f-1}\times\mathbb{CP}^{n_f-1}\times\cdots\times\mathbb{CP}^{n_f-1}}_{N\text{ copies}},
\end{align}
and using Dirac notation, a point in this phase space can be represented as 
\begin{align}
  |\phi_1\rangle \otimes |\phi_2\rangle \otimes \cdots \otimes|\phi_N\rangle, 
  \quad \text{with each }|\phi_i\rangle \in \mathbb{CP}^{n_f-1}.
\end{align}

Below, we first review how to parametrize $\mathbb{CP}^{n_f-1}$ in terms of homogeneous coordinates and how to impose the necessary constraints to remove the redundant phase. We then derive the induced Poisson (Dirac) brackets on $\mathbb{CP}^{n_f-1}$ and show how the classical spin observables satisfy the expected $\su(n_f)$ Poisson relations. Finally, we derive the classical equations of motion for an $\SU(n_f)$ spin Hamiltonian, and use Legendre transform to obtain the corresponding Lagrangian.

\subsection{Canonical Poisson (Dirac) brackets}\label{subsec:PB}

Let $\phi^\alpha: \alpha = 1,\cdots,n_f$ be the homogeneous coordinates in $\mathbb{C}^{n_f}$. We may regard these as being related to canonical real coordinates $q^\alpha, p^\alpha$ through
\begin{align}
  \phi^\alpha = \frac{1}{\sqrt{2}}(q^\alpha + \i p^\alpha),
  \quad
  \phistar^{\alpha} = \frac{1}{\sqrt{2}}(q^\alpha - \i p^\alpha).
\end{align}
In the usual canonical Poisson bracket\footnote{Note that these Poisson brackets corresponds to the standard symplectic form $\sum_i \d p_i \wedge \d q_i$ on the phase space $\mathbb{R}^{2n_f}$. This is equivalent to the symplectic form $\sum_i \i \d \phi_i^* \wedge \d \phi_i$ on $\mathbb{C}^{n_f}$ upon a change of coordinate, which gives rise to the Fubini-Study form when restricted to $\mathbb{CP}^{n_f-1}$.}, 
\begin{align}
  \{q^\alpha, p^\beta\}_{\mathrm{PB}} = \delta^{\alpha \beta}, \quad \{q^\alpha, q^\beta\}_{\mathrm{PB}}=0, \quad   \{p^\alpha, p^\beta\}_{\mathrm{PB}}=0,
\end{align}
one checks that
\begin{align}
  \{\phistar^{\alpha},\phi^\beta\}_{\mathrm{PB}} = \i\delta^{\alpha\beta}.
\end{align}
Equivalently, in bra-ket notation, setting $|\phi\rangle = \sum_\alpha \phi^\alpha|\alpha\>$ and $\langle\phi|=\sum_\alpha \<\alpha|\phistar^{\alpha}$, one can write
\begin{align}
  \{\<\phi|, |\phi\>\}_{\mathrm{PB}} = \i \mathbbm{1},
\end{align}

One may argue that the physical spin degree of freedom is represented by the projective space $\mathbb{CP}^{n_f - 1}$ rather than $\mathbb{C}^{n_f}$, and that Dirac bracket should be used rather than the Poisson bracket,
\begin{align}
  \{\<\phi|, |\phi\>\}_{\mathrm{DB}} = \i (\mathbbm{1} - |\phi\> \<\phi|).
\end{align}
However, as we will see in the next subsection, since the spin variables are manifestly invariant under $|\phi\> \mapsto e^{\i\theta}|\phi\>$, $|\phi\>$ is only subject to a single constraint $\<\phi|\phi\> = 1$, and the Dirac bracket is not necessary.

\subsection{Classical spin variables and their Poisson (Dirac) brackets}

The classical spin variables are the expectation values of the $\su(n_f)$ generators $T^a$ inside the coherent states,
\begin{align}
  S^a := \langle\phi| T^a|\phi\rangle, \quad a=1,2,\dots,n_f^2-1.
\end{align}
We wish to compute the Poisson bracket $\{S^a,S^b\}_{\mathrm{PB}}$. Using the known Poisson bracket of $\{|\phi\rangle,\langle\phi|\}_{\mathrm{PB}}$, one finds:
\begin{align}
  \{S^a,S^b\}_{\mathrm{PB}} &= \{\langle\phi|T^a|\phi\rangle,\langle\phi|T^b|\phi\rangle\}_{\mathrm{PB}} \nonumber\\
  &= \langle\phi|T^b\{\langle\phi|,|\phi\rangle\}_{\mathrm{PB}}T^a|\phi\rangle + \langle\phi|T^a\{|\phi\rangle,\langle\phi|\}_{\mathrm{PB}}T^b|\phi\rangle.
\end{align}
Since $\{\langle\phi|, |\phi\rangle\}_{\mathrm{PB}} = \i\mathbbm{1}$, we obtain
\begin{align}
  \{S^a,S^b\}_{\mathrm{PB}} &= -\i\langle\phi|[T^a,T^b]|\phi\rangle.
\end{align}
In $\mathfrak{su}(n_f)$, we have $[T^a,T^b] = \i \sum_cf^{abc}T^c$, so
\begin{align}
  \{S^a,S^b\}_{\mathrm{PB}} = \sum_cf^{abc}S^c.
\end{align}
which is precisely the expected $\mathfrak{su}(n_f)$ Poisson structure for the classical spin variables $S^a$. One can easily check that $\{S^a,S^b\}_{\mathrm{DB}} = \{S^a,S^b\}_{\mathrm{PB}}$.

\subsection{Hamiltonian Dynamics and Equations of Motion}

Consider now a classical spin Hamiltonian of the form
\begin{align}
  H = \sum_{i<j, a} J_{ij} \<\phi_i| T^a |\phi_i\> \<\phi_j| T^a |\phi_j\> = \sum_{i<j} J_{ij} S_i^aS_j^a.
\end{align}
Here, $|\phi_i\rangle\in\mathbb{CP}^{n_f-1}$ is the spin coherent state on site $i$, and $S_i^a = \langle \phi_i|T^a|\phi_i\rangle$. We want to find the time evolution of $|\phi_i(t)\rangle$ under the Hamiltonian flow associated with $H$. 

Since
\begin{align}\label{eq:psi-EOM-PB}
  \frac{\d}{\d t}|\phi_i\rangle = \{|\phi_i\rangle,H\}_{\mathrm{PB}},
\end{align}
we compute
\begin{align}
  \frac{\d}{\d t} |\phi_i\> &= \{|\phi_i\>, \sum_{j<k,a} J_{jk} \<\phi_j| T^a |\phi_j\> \<\phi_k| T^a |\phi_k\>\}_{\mathrm{PB}} \nonumber\\
  &= \sum_{j<k,a} J_{jk} \Big(\{|\phi_i\>, \<\phi_j|\}_{\mathrm{PB}}  T^a |\phi_j\> \<\phi_k| T^a |\phi_k\> + \<\phi_j| T^a |\phi_j\> \{|\phi_i\>, \<\phi_k|\}_{\mathrm{PB}} T^a |\phi_k\> \Big) \nonumber\\
  &= -\i \sum_{j,a} J_{ij} \<\phi_j| T^a |\phi_j\> T^a |\phi_i\> .
\end{align}
Thus $|\phi_i\rangle$ effectively evolves under a Schr\"odinger-like equation of coupled single-spin 
Hamiltonians
\begin{align}\label{eq:psi-Hi}
  \i \frac{\d}{\d t}|\phi_i\rangle = H_i|\phi_i\rangle \quad \text{ with } \quad H_i &= \sum_j J_{ij} \<\phi_j| T^a |\phi_j\> T^a = \sum_j J_{ij} S_j^a T^a .
\end{align}
Here we want to emphasize that since $|\phi_i\rangle$ also appear in the Hamiltonian, this equation is \emph{non-linear}.

If we instead using Dirac bracket in \cref{eq:psi-EOM-PB}, namely
\begin{align}
  \frac{\d}{\d t}|\phi_i\rangle = \{|\phi_i\rangle,H\}_{\mathrm{DB}},
\end{align}
then the effective single-spin Hamiltonian becomes
\begin{align}
  H_i^{\mathrm{DB}} &= \sum_j J_{ij} S_j^a (T^a - S_i^a) .
\end{align}
Compared to \cref{eq:psi-Hi}, the extra term is just a constant shift to the Hamiltonian, which indeed does not change the dynamics, as we argued in \cref{subsec:PB}. The effect of this extra constant is just to guarantee $\langle\phi_i|H_i^{\mathrm{DB}}|\phi_i\rangle = 0$, i.e. $H_i^{\mathrm{DB}}|\phi_i\rangle$ is always orthogonal to $|\phi_i\rangle$, ensuring it never leave $\mathbb{CP}^{n_f-1}$.

\subsection{Lagrangian of the classical spin model}\label{app:L}
Given the Hamiltonian, one can derive its Lagrangian using Legendre transform
\begin{align}
  L = \sum_{i,\alpha} p_i^\alpha\dot{q}_i^\alpha - H.
\end{align}
In terms of the $\phi_i^\alpha$ fields, one can write
\begin{align}
  \sum_i p_i^\alpha\dot{q}_i^\alpha &= \frac{1}{2\i}\sum_{i,\alpha} (\phi_i^\alpha-\phi_i^{\alpha*}) (\dot{\phi}_i^\alpha+\dot{\phi}_i^{\alpha*}) \nonumber\\
  &= \frac{1}{2\i}\sum_{i,\alpha} \phi_i^\alpha\dot{\phi}_i^\alpha+ \phi_i^\alpha\dot{\phi}_i^{\alpha*} -\phi_i^{\alpha*}\dot{\phi}_i^\alpha -\phi_i^{\alpha*}\dot{\phi}_i^{\alpha*}.
\end{align}
Since both $\phi_i^\alpha\dot{\phi}_i^\alpha$ and $\phi_i^{\alpha*}\dot{\phi}_i^{\alpha*}$ are total derivatives, they can be removed without affecting the dynamics. Therefore we have the Lagrangian
\begin{align}
  L = \frac{1}{2\i}\sum_{i,\alpha} (\phi_i^\alpha\dot{\phi}_i^{\alpha*} -\phi_i^{\alpha*}\dot{\phi}_i^\alpha) - \sum_{i<j} J_{ij} \<\phi_i| T^a |\phi_i\> \<\phi_j| T^a |\phi_j\>.
\end{align}
Equivalently, we can also subtract a total derivative $\frac{1}{2\i}\sum_i \frac{\d}{\d t}(\phi_i^\alpha\phi_i^{\alpha*})$, and write
\begin{align}
  L = \sum_i \langle\phi_i|\i\frac{\d}{\d t}|\phi_i\rangle - \sum_{i<j} J_{ij} \<\phi_i| T^a |\phi_i\> \<\phi_j| T^a |\phi_j\>,
\end{align}
which is identical to what one would obtain using coherent state path integral.

It can be checked that both forms of the Lagrangian lead to the same equation of motion for $|\phi_i\rangle$, namely \cref{eq:psi-Hi}, as expected.
\end{document}